\documentclass[12pt,twoside]{article}
\usepackage{latexsym,amsmath,amssymb,hyperref,mdframed,xcolor,graphicx}
\newdimen\paravsp  \paravsp=1.3ex
\usepackage{tikz}   

\definecolor{dc-red}{RGB}{224,1,77}
\definecolor{dc-magenta}{RGB}{138,21,56}

\hypersetup{
		breaklinks=true, 
		colorlinks = true,
		linkcolor={red},
		urlcolor={blue},
		citecolor={blue}
		}

\mdfdefinestyle{RedBox}{%
    linewidth=2pt,
    roundcorner=10mm,
    innertopmargin=5mm,
    innerbottommargin=5mm,
    innerrightmargin=10mm,
    innerleftmargin=10mm,
    frametitlerule=true,
    frametitlefontcolor=white,
    frametitlebackgroundcolor=dc-magenta
  }

\mdfsetup{nobreak=true} 

\AtBeginShipout{\AtBeginShipoutUpperLeft{
    \begin{tikzpicture}[remember picture,overlay]
      \node[anchor=north east,inner sep=0pt] at (current page.north east)
                 {
                  \includegraphics[scale=0.10]{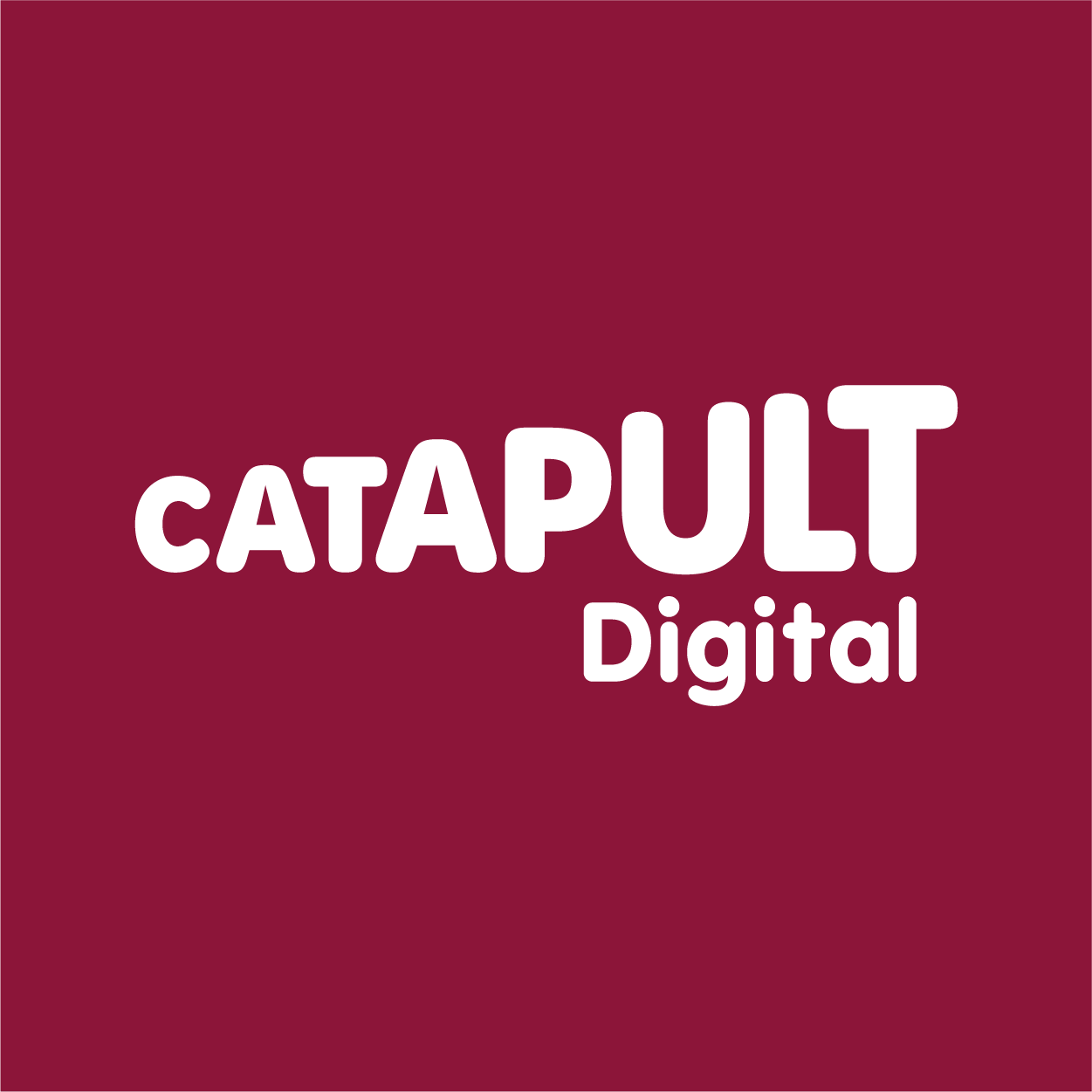}
                 };
    \end{tikzpicture}
}}

\begin{document}

\title{

\vskip 2mm\bf\Large\hrule height5pt \vskip 4mm
A Minimal Introduction to Quantum Computing
\vskip 4mm \hrule height2pt
}
\author{ 
  M M Hassan Mahmud \thanks{email: \href{mailto:hassan.mahmud@digicatapult.org.uk}{hassan.mahmud@digicatapult.org.uk}}
\and 
Daniel Goldsmith \thanks{email: \href{mailto:daniel.goldsmith@digicatapult.org.uk}{daniel.goldsmith@digicatapult.org.uk}} 
}

\date{Digital Catapult \\ [3mm]
\normalsize February 2025}
\maketitle
\begin{tikzpicture}[remember picture,overlay]
   \node[anchor=north east,inner sep=0pt] at (current page.north east)
              {\includegraphics[scale=0.1]{DC_Logo_Housed_Dark_Red.png}};
\end{tikzpicture}

\begin{abstract}
In this article, we present an introduction to quantum computing (QC) tailored for computing professionals such as programmers, machine learning engineers, and data scientists. Our approach abstracts away the physics underlying QC, which can be challenging to grasp, and frames it as a model of computation similar to, for instance, Turing machines. This helps readers grasp the fundamental principles of QC from a purely logical perspective. We begin by defining quantum states and qubits, establishing their mathematical representation and role in computation. We introduce fundamental concepts such as basis states, quantum gates, and tensor products, illustrating how these form the building blocks of quantum computation. Then we present the Deutsch–Jozsa algorithm, one of the simplest quantum algorithms that demonstrate how quantum computers can outperform classical computers. Finally, we provide guidance for further study, recommending resources for those interested in exploring quantum algorithms, implementations, and industry applications. 
\end{abstract}

\section{Introduction}

The goal of this article is to help ease understanding of quantum computing (QC) by computing professionals (for instance, if you are a programmer, or a machine learning engineer or a data scientist). This draws directly from the experience of the first author, who began this journey not too long ago as an AI/ML professional, but had the help of his colleague (the second author, Quantum Computing professional) to smooth the way.

Quantum computing can be a fairly challenging subject to understand even for people with a background in computing. A big reason for this is that it requires coming to grips with 

\begin{itemize}
\item concepts from quantum physics that are at odds with our day to day experience in the world, 
\item how these ideas are represented mathematically, 
\item how these mathematical notions can be used to carry out meaningful computation.
\end{itemize}

However, it turns out that there is a way to explain the fundamental ideas from quantum computing from a purely "logical" perspective -  this is the approach we take in this article. This abstracts away the physics behind QC completely, and looks at it as “just” another \href{https://en.wikipedia.org/wiki/Model_of_computation}{model of computation} \footnotemark[1]. 

\footnotetext[1]{In this article we adopt a hybrid approach of citing references because the intended mode of dissemination of this article is as an online article. For well known notions from computing, mathematics or quantum physics, we just use an in text hyperlink to an online resource (often Wikipedia). For more recent or research topics, we include proper citations.}

Examples of models of computation are \href{https://en.wikipedia.org/wiki/Turing_machine}{Turing machines} (the most well known), \href{https://en.wikipedia.org/wiki/Lambda_calculus}{lambda calculus}, \href{https://en.wikipedia.org/wiki/Combinatory_logic}{combinatory logic} or even your favourite programming language (with some caveats not important for this article). For our purposes, a model of computation is given by a set of symbols, and a set of rules for manipulating those symbols to carry out meaningful computations 

As an example, in a Turing machine, the symbols are the symbols on a tape, and the rules define how the machine moves between different states after inspecting the current symbol. In a programming language, the symbols constitute the language itself, and the rules define how a program is executed by the interpreter or the machine. 

When presented from this "logical" or model-of-computation perspective, understanding QC still poses some challenges. But we believe that this is manageable to do in 10 to 15 pages. Of course, to understand how QC is implemented using quantum computers, and to design effective quantum algorithms, understanding the physics behind is necessary to a certain extent. We believe that the logical perspective helps in understanding the full physics based picture, and also motivates the effort required in understanding it.

In the following sections, we first define what the set of symbols are in the context of QC, and how those symbols are manipulated to carry out computations. We then describe some quantum algorithms in that context at a high level and end with how the reader can continue their efforts in understanding and mastering quantum computing.

\subsection{Formal Models of Quantum Computation}

For the interested reader, there are well established formal models of quantum computing such as quantum logic gates \cite{AY1993} and quantum Turing machines \cite{PB1980}, and newer models with a much stronger connection to formal logic, such as quantum lambda calculus \cite{VT2004} and quantum typed lambda calculus \cite{SV2013}. In this article, we present a minimal introduction to the model corresponding to quantum logic gates while leaving out the physics entirely.

\section{Quantum states}

The ‘symbolic’ part of QC is represented by qubits - which are to QC what bits are to classical computing. Qubits in turn are defined by quantum states - and a  quantum state is represented as a weighted sum of basis states. To summarize 

\begin{quote}
Symbolic part of QC = qubits $\rightarrow_{\text{defined by}}$ quantum states $\rightarrow_{\text{defined by}}$ sum of basis states 
\end{quote}

\subsection{Basis States}

Basis states are the fundamental 'symbolic' units of quantum computing. Each basis state is represented by a binary string enclosed within $|\cdot\rangle$ which is called a {\em ket}. A related notation is $\langle\cdot|$ - this is called the {\em bra} which we discuss in more detail below. Together they are called the {\em bra-ket} notation. The following are some examples of basis states: 

\begin{mdframed}[style=RedBox, frametitle={{\bf Examples: Basis states}}]
\begin{enumerate}
  \item $|0\rangle$ is a basis state with the binary string $0$.
  \item $|1\rangle$ is a basis state with the binary string $1$.
  \item $|010010\rangle$ is a basis state with the binary string $010010$.
  \item $|001\rangle$ is a basis state with the binary string $001$.
\end{enumerate}
\end{mdframed}

Note that the number of basis states with binary strings of length $n$ is $2^n$ (corresponding to the number of binary strings of length $n$). The set of all basis states of the same length $n$ are representations of the \href{https://en.wikipedia.org/wiki/Orthonormal_basis}{orthonormal basis vectors}  in a $2^n$ dimensional \href{https://en.wikipedia.org/wiki/Vector_space}{vector space}. With this interpretation, $|\cdot\rangle$ denotes a column vector and $\langle\cdot|$ denotes a row vector.  Here are some examples:

\begin{mdframed}[style=RedBox, frametitle={{\bf Example: Basis states / kets as column vectors}}]

$|001\rangle$ is a \href{https://en.wikipedia.org/wiki/Standard_basis}{canonical} \href{https://en.wikipedia.org/wiki/Unit_vector}{unit} \href{https://en.wikipedia.org/wiki/Row_and_column_vectors}{column vector} in a $2^3$ dimensional space.
So $|001\rangle$ is the column vector of length $8$, with all $0$s except for row $1$ (because $001$ is binary representation of the integer $1$), with the row/column numbering starting at $0$
\begin{equation}
  \begin{bmatrix}
    0\\
    1\\
    0\\
    0\\
    0\\
    0\\
    0\\
    0\\
  \end{bmatrix}
  \end{equation}
The above generalises to arbitrary dimension $n$ (instead of $3$) in the obvious way.   
\end{mdframed}

\begin{mdframed}[style=RedBox, frametitle={{\bf Example: Bras as row vectors}}]
$\langle 001|$ is a canonical unit \href{https://en.wikipedia.org/wiki/Row_and_column_vectors}{row vector} in a $2^3$ dimensional space.
So $\langle 001|$ is the row vector of length $8$ with all 0s except for column $1$ (because $001$ is binary representation of the integer $1$), with the row/column numbering starting at $0$
\begin{equation}
  \begin{bmatrix}
    0 & 1 & 0 & 0 & 0 & 0 & 0 & 0
  \end{bmatrix}
  \end{equation}

The above generalises to arbitrary dimension $n$ (instead of $3$) in the obvious way. 
\end{mdframed}

Those familiar with standard vector notation may be annoyed at first with the bra-ket notation, but its utility becomes clearer as one works with it - you might even start using it outside Quantum computing!

\subsection{From Basis States to Quantum States}\label{from_basis_to_quant}

A quantum state is a weighted sum of basis states of the same dimension - that is, it is a vector in the space spanned by a set of basis vectors. 
\begin{mdframed}[style=RedBox, frametitle={{\bf Examples: Quantum States}}]
\begin{enumerate}
\item $|0 \rangle$
\item $i\frac{1}{\sqrt{2}}|01\rangle  + \frac{1}{\sqrt{2}}|11\rangle $ (‘$i$’ is the imaginary number) 
\item $i\frac{1}{2}|010010\rangle   + i \frac{1}{2}|110010\rangle   + \frac{1}{2}|011010\rangle   + \frac{1}{2}|010011\rangle $  (‘$i$’ is the imaginary number)
\end{enumerate}
\end{mdframed}

The weights need to satisfy the following requirements (which the examples satisfy, and you should check):

\begin{enumerate}
\item The weights are \href{https://en.wikipedia.org/wiki/Complex_number}{complex numbers}. 
\item The sum of the square of the \href{https://en.wikipedia.org/wiki/Absolute_value}{modulus} of the weights has to sum to 1. 
\end{enumerate}

The second property will play a big role when reading off the results of a quantum computation. A quantum state like $i\frac{1}{\sqrt{2}}|01\rangle  + \frac{1}{\sqrt{2}}|11\rangle $ is said to be in a {\bf superposition} of  the basis states $|01\rangle $ and $|11\rangle $. 

\begin{quote}
If you are wondering about the 'why' of it all at this point, we ask you to remember what we said at the beginning of this article: all this follows from theory of quantum physics and the experimental results which support it. In this article we are ignoring the physics, and just presenting it as a 'logical system' of symbols and rules for manipulating those symbols.  
\end{quote}

\subsection{A Notational Simplification}

For convenience, a basis state $|x\rangle $ is written as $|n_x\rangle $,  where $n_x$ is the number corresponding to the binary string $x$. 
\begin{mdframed}[style=RedBox, frametitle={{\bf Examples: Notational simplificiation}}]
\begin{enumerate}
\item $|000001\rangle $ is denoted by $|1\rangle $ (the decimal representation of $000001$). 
\item $|100001\rangle $ is denoted by $|33\rangle $ (the decimal representation of $100001$).
\item $|01\rangle $ is (also) denoted by $|1\rangle $ (the decimal representation of $01$).
\end{enumerate}
\end{mdframed}

That is, we replace the binary string by its corresponding decimal representation. Note that the (1) and (3) above have the same denotation. This means that when we use this notation, it must be clear from the context which set of basis states we are using. Using this notation, the example quantum states above are written as:

\begin{mdframed}[style=RedBox, frametitle={{\bf Examples: Quantum state - simplified notation}}]
  \begin{enumerate}
    \item $|0\rangle$ is a quantum state over $2^1$ qubits.  
    \item $i\frac{1}{\sqrt{2}}|1\rangle  + \frac{1}{\sqrt{2}}|3\rangle$ is a quantum state over $2^2$ qubits.
    \item $i\frac{1}{2}|18\rangle   + i\frac{1}{2}|50\rangle   + \frac{1}{2}|26\rangle   + \frac{1}{2}|19\rangle$  is a quantum state over $2^8$ qubits. 
    \end{enumerate}
  \end{mdframed}

\subsection{Dot and Outer Products} 

In the following we define the dot and outer products for the basis states, and then quantum states in general. These are vital notions for defining quantum computation later on. 

\subsubsection{Dot and Outer Products for Basis States}\label{sec_dot_and_outer}

The \href{https://en.wikipedia.org/wiki/Dot_product}{dot product} $| y \rangle \cdot | x \rangle$ between two basis states  of the same dimension is written as  $\langle x|y\rangle$.  

\begin{quote}
Note that since $|x\rangle$ and $|y\rangle$ are orthonormal basis vectors, 
$\langle x|y \rangle  = 1$ if $x = y$ (i.e. the same binary string) and $\langle x|y\rangle  = 0$ if $x \neq y$ (i.e. different binary strings).
\end{quote}

Similarly, the \href{https://en.wikipedia.org/wiki/Outer_product}{outer product} is written as $|x\rangle \langle y|$. As per the definition, the outer product of two vectors in an $m$ dimensional space gives an $m \times m$  matrix.

\begin{quote}
In particular because $|x\rangle$ and $\langle y|$ are unit vectors, and interpreting $x$ and $y$ as binary representation of numbers, in the $n\times n$ matrix $|x\rangle \langle y|$, the entry in row $i$ and column $j$ $|x\rangle \langle y|_{i, j} = 1 $ if $i = x$ and $j = y$, and otherwise $|x\rangle \langle y|_{i, j} = 0$. Here we are indexing the matrix rows and columns from $0$ to $m-1$. This follows directly from the definition of outer products. For example, if $x = |1000\rangle$ and $y = \langle 0001|$, then $i = 8$ and $j = 1$, and in the $4 \times 4$ matrix $|x\rangle \langle y|$, the entry $|x\rangle \langle y|_{8, 1} = 1$, and all the other entries are 0.
\end{quote}

As we are working with vectors, the products of these follow the usual rules for linear algebra. In particular, by associativity, we have 

\begin{equation}\nonumber
\left (|w\rangle \langle x| \right)|y\rangle  = |w\rangle \left(\langle x|y\rangle \right)
\end{equation}

The above has the interesting consequence that for basis states, the dot product $\langle x|y\rangle $ can be thought of as a {\bf selection tool}. 
\begin{quote}
if $x = y$ then $(|w\rangle \langle x|)|y\rangle  = |w\rangle (\langle x|y\rangle ) = |w\rangle 1 = |w\rangle $ and 
if $x \neq y$, $(|w\rangle \langle x|)|y\rangle  = |w\rangle (\langle x|y\rangle ) = |w\rangle 0 = \mathbf{0}$ (where $\mathbf{0}$ is the zero vector). 
\end{quote}
We will use the above fact to define operators over quantum states later on.

\subsubsection{Dot and Outer Products for Quantum States}

Dot and outer products between quantum states is extended to quantum states in the usual way by distributing the dot product of the basis states that make up the quantum states. However, now we also need to deal with the weights of each basis state making up the quantum state. 

For dot products, we take the \href{}{complex conjugate} of the weight of the second vector and then do a complex multiplication of the weights. An example will make it clear. For instance, say,
\begin{align}
\nonumber \phi &= i\frac{1}{\sqrt{2}}|1\rangle  + \frac{1}{\sqrt{2}}|3\rangle \\
\nonumber \psi &= i\frac{1}{\sqrt{3}}|1\rangle  + \frac{1}{\sqrt{3}}|2\rangle + \frac{1}{\sqrt{3}}|3\rangle
\end{align}
then 
\begin{align}
\nonumber  | \psi \rangle | \phi \rangle = \langle \phi | \psi \rangle = &  (-i)\frac{1}{\sqrt{2}}\cdot i\frac{1}{\sqrt{3}}  \langle 1 |1\rangle + (-i)\frac{1}{\sqrt{2}}\cdot i\frac{1}{\sqrt{3}}  \langle 1 |2 \rangle + (-i)\frac{1}{\sqrt{2}}\cdot \frac{1}{\sqrt{3}}  \langle 1 |3\rangle  \\
 \nonumber & + \frac{1}{\sqrt{2}}\cdot i\frac{1}{\sqrt{3}}  \langle 3 |1\rangle +  \frac{1}{\sqrt{2}}\cdot i\frac{1}{\sqrt{3}}  \langle 3|2 \rangle +  \frac{1}{\sqrt{2}}\cdot i\frac{1}{\sqrt{3}}  \langle 3 |3\rangle\\
\nonumber = &\frac{1}{\sqrt{6}} + \frac{1}{\sqrt{6}} = \frac{\sqrt{2}}{\sqrt{3}}
\end{align}
  
First note that all the terms $\langle i |j \rangle$ where $i \neq j$ vanish as per the discussion above. So, in the second line, 
the first $\frac{1}{\sqrt{6}}$ comes from the weights for the basis state $|1\rangle$ in the two terms: $i\frac{1}{\sqrt{2}}$. The product after taking the complex conjugate is $i\frac{1}{\sqrt{3}} \times (-i\frac{1}{\sqrt{2}})$ = $\frac{1}{\sqrt{6}}$. The second $\sqrt{6}$ comes from the weights for the basis state $|3\rangle$ (not shown in the first line) and the product in this case is $i\frac{1}{\sqrt{2}} \times \frac{1}{\sqrt{2}}  = \frac{\sqrt{2}}{\sqrt{3}}$

The outer product is defined in a similar distributive manner: 

\begin{align}
  \nonumber | \psi \rangle \langle \phi | = &  i\frac{1}{\sqrt{3}} \cdot (-i)\frac{1}{\sqrt{2}}  |1\rangle \langle 1| + i\frac{1}{\sqrt{3}} \cdot \frac{1}{\sqrt{2}}  |1\rangle \langle 3| \\
  \nonumber  & + \frac{1}{\sqrt{3}} \cdot (-i)\frac{1}{\sqrt{2}}  |2\rangle \langle 1| + \frac{1}{\sqrt{3}} \cdot \frac{1}{\sqrt{2}}  | 2 \rangle \langle 3| \\
  \nonumber  & + \frac{1}{\sqrt{3}} \cdot (-i)\frac{1}{\sqrt{2}}  |3\rangle \langle 1| + \frac{1}{\sqrt{3}} \cdot \frac{1}{\sqrt{2}}  | 3 \rangle \langle 3| \\
  \nonumber = & \frac{1}{\sqrt{6}} |1\rangle \langle 1| + i\frac{1}{\sqrt{6}}  |1\rangle \langle 3| \\
  \nonumber  &  - i\frac{1}{\sqrt{6}} |2\rangle \langle 1| + \frac{1}{\sqrt{6}}  | 2 \rangle \langle 3| \\
  \nonumber  &  - i\frac{1}{\sqrt{6}} |3\rangle \langle 1| + \frac{1}{\sqrt{6}}   | 3 \rangle \langle 3| 
\end{align}

\subsection{Qubits}

Qubit refers to the length of the basis states that comprise a quantum state. 

\begin{mdframed}[style=RedBox, frametitle={{\bf Examples: Qubits}}]
\begin{enumerate}
\item Each of the quantum states $|0\rangle $, $|1\rangle $ consists of 1 qubit.
\item Each of the quantum states $|10\rangle $, $|11\rangle $ consists of 2 qubits.
\item The quantum state $i\frac{1}{2}|010010\rangle   + i\frac{1}{2}|110010\rangle   + \frac{1}{2}|011010\rangle   + \frac{1}{2}|010011\rangle $ consists of $6$ qubits. 
\end{enumerate}
\end{mdframed}

This concludes our presentation of the 'symbols' part of quantum computation. We will now look at how these symbols are used for computing useful things.

\section{ Computation in Quantum Computing}

In this section we look at how qubits are manipulated for computation and how the corresponding results are read off for output to human users. As we will see, the output of a given run of a quantum computation is a set of classical bits like the ones we are used to in normal computation. However, for multiple runs for the same input, the outputs are probabilistic - this is one of the unique aspects of QC. 

\subsection{Quantum Gates}

{\bf Quantum gates} or {\bf quantum logic gates} or {\bf quantum operators} are the main computational units in quantum computing. They transform one quantum state into another. A quantum operator $O$ that can be applied to an n-qubit state is written as a weighted, double sum over the set of matrices formed from the outer product of the basis states in $2^n$ dimensions:

\begin{equation}\nonumber
O = \sum_{i=0}^{2^{n}-1}\sum_{j=0}^{2^{n}-1} a_{i, j} |i\rangle \langle j|    
\end{equation}

Recalling the selection tool idea from Section \ref{sec_dot_and_outer}, the $O$ is a matrix or a linear operator in the space defined by the basis states, and the sum on the right is merely setting $O_{i, j}$ to $a_{i, j}$. What makes two quantum operators distinct from each other is having different values for $a_{i,j}$. Note that the weights $a_{i, j}$ are complex valued and can be $0$.

\subsubsection{Some Commonly Used Quantum Gates}\label{sec_gate_examples}

Here are some examples of commonly used quantum logic gates. It is worth working out the operation of the gates on single qubits using the selection tool idea that was mentioned above (this should be trivial - but we will show what these do below anyway).

\noindent {\em Identity Gate}: Denoted by $I$, and defined as 
\begin{equation}\nonumber
I = | 0 \rangle \langle 0| + | 1 \rangle \langle 1| 
\end{equation}

\noindent {\em Not Gate}: Denoted by $X$ and defined as 
\begin{equation}\nonumber
X  = | 0 \rangle \langle 1|  + | 1 \rangle \langle 0| 
\end{equation}

\noindent {\em Hadamard Transformation}: Denoted by $H$, this is a commonly used gate that puts a single qubit into a superposition of states. This is defined by  
\begin{equation}\nonumber
H = \frac{1}{\sqrt{2}} \left ( | 0 \rangle \langle 0|  + | 1 \rangle \langle 0| + | 0 \rangle \langle 1|  - | 1 \rangle \langle 1 | \right )
\end{equation}

\noindent {\em Controlled-NOT gate}: Denoted by $C_{not}$ it is a commonly used gate and it operates on two qubits. This is defined by:

\begin{equation}\nonumber
C_{not} = | 00 \rangle \langle 00|  + | 01 \rangle \langle 01| + | 11 \rangle \langle 10|  + | 10 \rangle \langle 11 | 
\end{equation}

\subsection{Quantum Gate Computations}

Given a quantum state $\psi = \sum_{k=0}^{2^{n}-1} b_k |k\rangle $  (where some of the $b_k$ can be $0$), $O(\psi)$ is computed in two steps.

\subsubsection{ Step One}

The first step in computing $O(\psi)$ is computing $O\psi$, the matrix product of $O$ and $\psi$. It is computed in the usual manner:

\begin{equation}\nonumber
O\psi =  \sum_{i=0}^{2^{n}-1}\sum_{j=0}^{2^{n}-1} \sum_{k=0}^{2^{n}-1} \big (a_{i, j}|i\rangle \langle j| \big ) \; \big (b_k |k\rangle \big )  
\end{equation}

From the discussion in Section \ref{sec_dot_and_outer}, each term in the above sum can be simplified as follows

\begin{equation}\nonumber
\big (a_{i, j}|i\rangle \langle j| \big) \; b_k |k\rangle = a_{i, j}b_k \; \big (|i\rangle \langle j|\big ) \; |k\rangle = a_{i, j}b_k |i\rangle  \;\big (\langle j|k\rangle \big ) = a_{i, j}b_k |i\rangle \delta_{j, k} 
\end{equation}

where $\delta_{j, k}$ is the Kroneckar-delta function. So terms in the triple sum vanish unless $j = k$, and if $j = k$ the vector part of the term is just $|i\rangle$. So the first step of application of the operator $O(\psi)$ reduces to: 

\begin{equation}\nonumber
O\psi = \sum_{i=0}^{2^{n}-1}\sum_{k=0}^{2^{n}-1} a_{i, k} b_k |i\rangle = \sum_{i=0}^{2^{n}-1} |i\rangle \sum_{k=0}^{2^{n}-1} a_{i, k} b_k 
\end{equation}

We can simplify it even further, and drop the terms for which either $a_{i, k}$ or $b_{i, k}$ is $0$:

\begin{equation}\nonumber
O\psi = \sum_{i=0}^{2^n-1} |i\rangle \sum_{a_{i, k}\neq 0, b_k \neq 0} a_{i, k} b_k
\end{equation}

So we get (unsurprisingly) an expression for $O\psi$ as a weighted sum of basis states with each basis state $|i\rangle$ having weight $\sum_{a_{i, k}\neq 0, b_k \neq 0} a_{i, k} b_k$.

\subsubsection{Step Two}

Now recall from section \ref{from_basis_to_quant} that the square of the modulus of the weights for a quantum state needs to be normalized. So we need a factor of $1/A_{O,\psi}$ for $O\psi$ to be a valid quantum state, where 

\begin{equation}\nonumber
A_{O,\psi} = \sqrt{\sum_i \left |\sum_{a_{i, k}\neq 0, b_k \neq 0}b_k a_{i, k} \right|^2}
\end{equation}
Here $|.|$ is the modulus of a complex number So now we have the second step of computing $O(\psi)$, which is the normalisation by $A_{O, \psi}$. This gives us our final definition:

\begin{equation}\nonumber
O(\psi) = \sum_{i=0}^{2^n-1} |i\rangle \frac{\sum_{a_{i, k}\neq 0, b_k \neq 0}b_k a_{i, k}}{A_{O,q}}
\end{equation}

\subsubsection{Computing With Some Commonly Used Quantum Gates}

Here we show what the gates mentioned in section \ref{sec_gate_examples} compute. 

\noindent {\em Identity Gate}: Recall that $I = | 0 \rangle \langle 0| + | 1 \rangle \langle 1|$. This leaves a quantum basis state unchanged 
\begin{align}
\nonumber I| 0 \rangle &= | 0 \rangle \\ 
\nonumber I| 1 \rangle &= | 1 \rangle
\end{align}

\noindent {\em Not Gate}: Recall that $X = | 0 \rangle \langle 1|  + | 1 \rangle \langle 0|$. It flips a basis state 
\begin{align}
\nonumber X| 0 \rangle &= | 1 \rangle \\
\nonumber X| 1 \rangle &= | 0 \rangle
\end{align}

\noindent {\em Hadamard Transformation}: Recall that
\begin{equation}\nonumber
H = \frac{1}{\sqrt{2}} \left ( | 0 \rangle \langle 0|  + | 1 \rangle \langle 0| + | 0 \rangle \langle 1|  - | 1 \rangle \langle 1 | \right )
\end{equation}

This transforms single qubits in the following ways:
\begin{equation}\nonumber
H|0\rangle = \frac{1}{\sqrt{2}} \left ( | 0 \rangle  + | 1 \rangle \right )
\end{equation}
and
\begin{equation}\nonumber
H|1\rangle = \frac{1}{\sqrt{2}} \left ( | 0 \rangle  - | 1 \rangle \right )
\end{equation}

So this puts the input state into a uniform superposition of the basis states.

{\em Controlled-NOT gate}: Denoted by $C_{not}$, it is a commonly used gate, and it operates on two qubits. This is defined by:

\begin{equation}\nonumber
    C_{not} = | 00 \rangle \langle 00|  + | 01 \rangle \langle 01| + | 11 \rangle \langle 10|  + | 10 \rangle \langle 11 | 
\end{equation}

This changes the two qubits in the following way:

\begin{align}
 \nonumber C_{not}| 00 \rangle & = | 00 \rangle \\
 \nonumber C_{not}| 01 \rangle & = | 01 \rangle \\
 \nonumber C_{not}| 10 \rangle & = | 11 \rangle \\
 \nonumber C_{not}| 11 \rangle & = | 10 \rangle
\end{align}

You can see why this is called a controlled-not: if the first qubit is $0$, then the second qubit is passed unchanged - however if the first qubit is $1$, the second qubit is flipped. This is gate is like a not gate, which is activated based on the value of the first qubit.

\subsubsection{Quantum Gates Are Unitary}

We now discuss a very important property every quantum gate $O$ needs to satisfy. So far we have taken this into account implicitly. This property is that $O$ has to be {\em unitary}, meaning that $OO^\dagger = O^\dagger O = I$ where $O^\dagger$ is the \href{https://en.wikipedia.org/wiki/Conjugate_transpose}{conjugate transpose} of $O$. This in turn means $O$ has the following properties
  
\begin{enumerate}
\item $|O\psi| = |\psi|$ i.e. it preserves length.
\item $O$ is invertible - that is if $O\psi = \phi$, then we can get back $\psi = A\phi$ ($A$ is trivially $O^\dagger$).
\end{enumerate}

Property 1 follows from the definition of length of vectors: 
  \begin{equation}\nonumber
     |\psi| = \sqrt{\langle\psi |\psi \rangle } = \sqrt{\psi^\dagger \psi}
  \end{equation} 
  For a unitary matrix this becomes 
  \begin{equation}\nonumber
     |O\psi| = \sqrt{\langle O\psi |O\psi \rangle } = (O\psi)^\dagger (O\psi) = \psi^\dagger O^\dagger O\psi =  \psi^\dagger I\psi = \psi^\dagger \psi
  \end{equation}

The need for quantum gates to be unitary follows from quantum mechanics directly and is a consequence of the requirement that the probabilities of the different quantum states need to add up to one, as described in the next section.  

\subsection{Reading Off Quantum States}

The nature of quantum physics entails that we can never directly access the Quantum states. In particular, we have to perform a special operation on a quantum state and at that point we get probabilistic readout. Remember that each quantum state $\psi$ is a superposition of the set of $n$-qubit basis states:

\begin{equation}\nonumber
\psi = \sum_{k=0}^{2^{n}-1} b_k |k\rangle 
\end{equation}

When a quantum state is read off, or {\bf measured}, the output is one of the basis states with non-zero weight $b_k$ chosen probabilistically. The probability of observing $|k\rangle$ is $|b_k|^2$ -- the numbers work out because the sum of square of the modulus of the weights of a quantum state are required to sum to $1$. Indeed, for this reason, the weights $b_k$ are called {\bf probability amplitudes}. Finally, one observes the basis state $|k\rangle$, what they 'see' is the binary string for $k$.
\begin{mdframed}[style=RedBox, frametitle={{\bf Examples: Measuring quantum states}}]
\begin{enumerate}
\item Measuring $|0\rangle $ one observes the bit $0$ deterministically.
\item Measuring $i\frac{1}{\sqrt{3}}|1\rangle  + \frac{1}{\sqrt{2/3}}|3\rangle $ one observes $01$ and $11$ with probabilities $1/3$ and $2/3$ respectively.
\item Measuring $i\frac{1}{2}|18\rangle   + i\frac{1}{2}|50\rangle   + \frac{1}{2}|26\rangle   + \frac{1}{2}|19\rangle $ one observes the bit strings $010010$, $110010$,  $011010$ and $010011$, each with probability $1/4$.
\end{enumerate}
\end{mdframed}

\subsection{Tensor Products}

We now present tensor products, which are a type of operations in linear algebra. Tensor products are used to combine $m$-qubit vectors and $n$-qubit vectors to get a $m+n$-qubit vector. They are also used to combine quantum gates to form the quantum circuits that define quantum algorithms. 

\subsubsection{Tensor Products of Quantum States}

We will start with an example with $2$ qubits: $\psi = a_1|0\rangle + a_2|1\rangle$ and $\phi = b_1|0\rangle + b_2|1\rangle$. The tensor product $\psi \otimes \phi$ is defined as:

\begin{align}
\nonumber
\psi \otimes \phi & =  
a_1b_1 |0\rangle \otimes |0\rangle +  a_1b_2|0\rangle \otimes |1\rangle + a_2b_1|1\rangle \otimes |0\rangle + a_2b_2|1\rangle \otimes |1\rangle\\
\nonumber & = a_1b_1 |00\rangle +  a_1b_2|01\rangle + a_2b_1|10\rangle + a_2b_2|11\rangle
\end{align}

That is, in a tensor product, the basis elements defining the two vectors distributes. Also note that $2$-qubit basis states like $|00\rangle$ are just tensor product of $1$-qubit basis states $|0\rangle \otimes |0\rangle$. So we have been using tensor products, but just did not know about it.

In terms of the more traditional column vector representation, the tensor product $|0\rangle \otimes |0\rangle = |00\rangle$ is just a different way of writing


\begin{equation}\nonumber
(1, 0)^T \otimes (1, 0)^T = (1, 0, 0, 0)^T
\end{equation}

where $^T$ is the transpose. This operation multiplies each coordinate of the first one by the second one (so a scalar $\times$ a vector) and then stacks them on top of each other. The really cool thing here is that the bra-ket notation makes it very easy to work out what the tensor product should be.

This generalizes to tensor products of $m$-qubit and $n$-qubit states in the obvious way to give us $m+n$-qubit states. Make special note of the following, which often trips up newcomers:

\begin{quote}
$m$-qubits are actually elements of a $2^m$ dimensional vector space. So the tensor product of $m$-qubit and $n$-qubits gives us elements of $2^{m+n}$ dimensional vector-space!
\end{quote}

The general definition of tensor products can be found \href{https://en.wikipedia.org/wiki/Tensor_product}{here}.

\subsubsection{Tensor Products of Quantum Logic Gates}

Tensor products of quantum logic gates have a simple definition as well - you simply concatenate the strings in the bras and kets. Here is simple example. Say $X_{simple} = |1 \rangle \langle 0|$ - then recalling that $X = | 0 \rangle \langle 1|  + | 1 \rangle \langle 0|$, we have

\begin{equation}\nonumber
X \otimes X_{simple} =  | 01 \rangle \langle 10|  + | 11 \rangle \langle 00|
\end{equation}

So we concatenate the $1$ in the definition of $X_{simple}$ to the first sub-term of each of the outer products in the definition of $X$ and the $0$ in the definition of $X_{simple}$ to the second term of each of the outer products in the definition of $X$.

Taking a more complex example, recalling that 

\begin{equation}\nonumber
H = \frac{1}{\sqrt{2}} \left ( | 0 \rangle \langle 0|  + | 1 \rangle \langle 0| + | 0 \rangle \langle 1|  - | 1 \rangle \langle 1 | \right )
\end{equation}

we have,

\begin{align}
\nonumber H \otimes X =& \frac{1}{\sqrt{2}} \big ( |00 \rangle \langle 01|  + | 10 \rangle \langle 01| + | 00 \rangle \langle 11|  - | 10 \rangle \langle 11 | \\
\nonumber  & \;\; + | 01 \rangle \langle 00|  + | 11 \rangle \langle 00| + | 01 \rangle \langle 10|  - | 11 \rangle \langle 10 | \big ) \\
\end{align}


In terms of the more traditional matrix of numbers representation, the tensor product $H \otimes X$ replaces each component $H_{i, j}$ with the matrix $H_{i, j} \times X$ (i.e. a scalar $\times$ a matrix). So the tensor product of an $m \times m$ matrix with an $n \times n$ matrix is an $mn \times mn$ matrix. Again, the beauty of the bra-ket notation is that it makes pen-and-paper calculation of tensor products a straighforward exercise.

\subsubsection{Entanglement}

The final topic we discuss in this section is entanglement, which is quite important for implementing quantum algorithms in practice. An $n$-qubit quantum state is said to be in an entangled state if it cannot be written as tensor product of $n$ single qubit states. For instance consider the state $4$-qubit state $\frac{1}{\sqrt{2}}|00\rangle + \frac{1}{\sqrt{2}}|11\rangle$ which cannot be represented as a tensor product of single qubit states. To see this, if this were not true, then we need 
\begin{equation}\nonumber
 (a_1|0\rangle + b_1|0\rangle) \otimes (a_2|0\rangle + b_2|0\rangle) = \frac{1}{\sqrt{2}}|00\rangle + \frac{1}{\sqrt{2}}|11\rangle
\end{equation}
But we have,
\begin{equation}\nonumber
(a_1|0\rangle + b_1|0\rangle) \otimes (a_2|0\rangle + b_2|0\rangle) = a_1a_2|00\rangle + a_1b_2|01\rangle b_1a_2|10\rangle + b_1b_2|11\rangle 
\end{equation}
Equating the above two, we now need $a_1b_2 = 0$, which means $a_1 = 0$ or $b_2 = 0$, and so we cannot get the qubit we above. Please refer to \cite{RP2014} additional details on entanglement. 

\section{Deutsch–Jozsa Algorithm: Your First Quantum Algorithm}

In this section we present one of the simplest quantum algorithms that shows their advantage over classical algorithms. While the problem is somewhat artificial, it takes the quantum algorithm constant time (a single step in fact), but it takes a classical algorithm time exponential in the size of the input. 

\subsection{The Problem}

The problem that is solved by the Deutsch–Jozsa algorithm is the following. We are given a function $f : \{0, 1\}^n \rightarrow \{0, 1\}$ - i.e. a function from $n$ bits to $1$ bit - which takes one of three forms:

\begin{enumerate}
\item $f(x) = 0$ for all $x \in \{0, 1\}^n$
\item $f(x) = 1$ for all $x \in \{0, 1\}^n$ 
\item $f(x) = 0$ for half the bits in $\{0, 1\}^n$ and $f(x) = 1$ for the other half.
\end{enumerate}

In the first two cases $f$ is called {\em constant} and in the last case $f$ is called {\em balanced}. The problem is to determine whether $f$ is constant or if $f$ is balanced.

\subsection{The Classical Solution}

There are two possible solutions in the classical setting - a deterministic algorithm and a probabilistic algorithm. The deterministic algorithm essentially consists of trying out every possible $n$-bit one at a time and check if $f(x) = 1$ or $f(x) = 0$. If the true case is 1 or 2 (i.e. $f$ is constant), it takes exponential time ($2^{n-1} + 1$ steps) at least to determine the answer. If the true case is 3 (i.e. $f$ is balanced) the worst case is still the same, but if the algorithm is lucky, in the sequence chosen, it may find it sooner. 

The probabilistic algorithm is to sample $n$-bits uniformly at random, and test this value. In this case, the algorithm can determine the true case with high probability (failing with probability $\epsilon < 2^{-k}$ after $k$ steps). But to be absolutely sure, the worst case run time is the same.

\subsection{The Quantum Solution: The Deutsch–Jozsa Algorithm}

We should warn, that the math gets a bit heavier in this section. 

\begin{figure}
  \centering
  \includegraphics[width=0.75\linewidth]{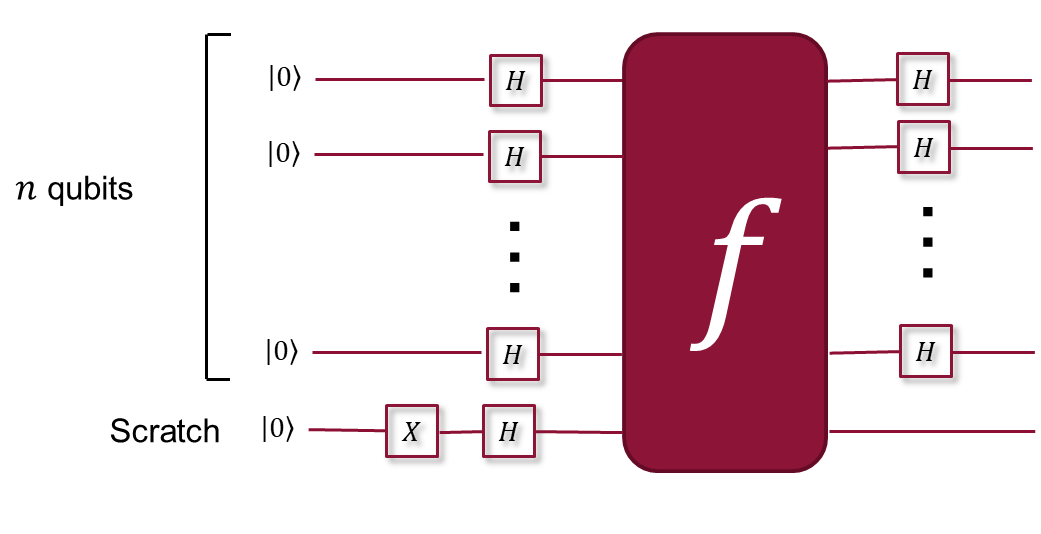}
  \caption{The quantum gate representation of the Deutsch-Josza algorithm.}\label{fig_DJ}
  \end{figure}

We follow the presentation found \href{https://en.wikipedia.org/wiki/Deutsch-Jozsa_algorithm}{here}. The quantum algorithm for solving this problem is defined by a quantum circuit that is made up of multiple quantum gates used in sequence. The Deustch-Josza algorithm is (see Figure \ref{fig_DJ}):

\begin{enumerate}
\item Apply the $n+1$-qubit Hadamard gate $H^{\otimes n+1}$ to $\phi \;_=^{def} \;|0\rangle^{\otimes n}|1\rangle$
\item Apply $U_f$, the quantum oracle for $f$ (see below), to $H^{\otimes n+1}(\phi)$.
\item Apply the $n$-qubit Hadarmard gate to the first $n$ qubit of the result of the previous step $U_f \circ H^{\otimes n+1}(\phi)$.
  \begin{itemize} 
  \item This means we apply $H^{\otimes n}I$ to $U_f \circ H^{\otimes n+1}(\phi)$, where $I$ is the identity gate applied to the last qubit.
  \end{itemize}
\item Measure the first $n$-qubits of $H^{\otimes n}I \circ U_f \circ H^{\otimes n+1}(\phi)$. If this measured state is $|0\rangle$, then $f$ is constant - otherwise it is balanced.
\end{enumerate}

So we can solve this problem in 3 applications of quantum gates, followed by a measurement. Therefore, we have an exponential speedup when comparing this to the time required by classical algorithms!

We now explain each step in detail and justify the algorithm.

\subsubsection{Step 1}

The algorithm starts with the $n+1$ qubit $|0\rangle^{\otimes n}|1\rangle$. Here $|0\rangle^{\otimes n}$ is the $n$ fold tensor product of the 1-qubit $|0\rangle$ and so $|0\rangle^{\otimes n}|1\rangle$ is the $n+1$ qubit $|1\rangle$) from the discussion above. 

The first step in the algorithm is to apply the 1-qubit Hadarmard gate to each of the $n+1$ qubits (this is also equivalent to applying the $n+1$ qubit version of the Hadamard gate to the $n+1$ qubits). This is written as $H^{\otimes n+1}$ and gives:

\begin{equation}\nonumber
H^{\otimes n+1}(|0\rangle^{\otimes n}|1\rangle) = \frac{1}{\sqrt{2^{n+1}}} \sum_{x=0}^{2^n-1} |x\rangle \otimes (|0\rangle - |1\rangle)
\end{equation}

Note that you can calculate the above by pen and paper by the rules around working with quantum gates we have identified in the previous sections 

\subsubsection{Step 2}

Next, the algorithm assumes that we have the function $f$ implemented as a {\em quantum oracle} $U_f$. This oracle maps $|x\rangle|y\rangle$ to $|x\rangle|y \oplus f(x)\rangle$ where $x$ is $n$-bits, $y$ is $1$-bits and $\oplus$ is the binary XOR operator. 

Given this oracle, after calculating the Hadamard transform, the Deutsch-Josza algorithm applies $U_f$ - which gives:

\begin{equation}\nonumber
U_f \circ H^{\otimes n+1}(|0\rangle^{\otimes n}|1\rangle) = \frac{1}{\sqrt{2^{n+1}}} \sum_{x=0}^{2^n-1} |x\rangle \otimes (|0 \oplus f(x) \rangle - |1 \oplus f(x) \rangle)
\end{equation}
In the l.h.s. of the above, $f(x)$ is either $0$ or $1$ for each $x$. 

So if $f(x) = 0$, we have 
\begin{equation}\nonumber
|0 \oplus f(x) \rangle - |1 \oplus f(x) \rangle = |0\rangle - |1\rangle
\end{equation}
whereas, if $f(x) = 1$, we have 
\begin{equation}\nonumber
|0 \oplus f(x) \rangle - |1 \oplus f(x) \rangle = |1\rangle - |0\rangle
\end{equation}

So we can simplify:

\begin{equation}\nonumber
U_f \circ H^{\otimes n+1}(|0\rangle^{\otimes n}|1\rangle) = \frac{1}{\sqrt{2^{n+1}}} \sum_{x=0}^{2^n-1} (-1)^{f(x)} |x\rangle \otimes (|0\rangle - |1\rangle)
\end{equation}

\subsubsection{Step 3}

In this step, the algorithm focuses only on the first $n$-qubits of output from the previous step. These qubits are:

\begin{equation}\nonumber
\frac{1}{\sqrt{2^{n}}} \sum_{x=0}^{2^n-1} (-1)^{f(x)} |x\rangle \; _=^{def} \; \psi
\end{equation}

Now we apply the $n$-qubit Hadamard transform $H^{\otimes n}$ to $\psi$ the above. An alternative way to write $n$-qubit Hadamard operator is: 

\begin{equation}\nonumber
H^{\otimes n}(k)  = \frac{1}{\sqrt{2^n}} \sum_{j=0}^{2^n-1} (-1)^{k\cdot j} |j\rangle
\end{equation}
where $k \cdot j$ is the addition modulo-2 of the bitwise products $=k_0j_0 \oplus k_1j_1\oplus ... \oplus  k_{n-1}j_{n-1}$.

Applying $H^{\otimes n}$ to $\psi$, by the linearity of quantum operators we have 

\begin{align}\nonumber
H^{\otimes n}(\psi)  &= \frac{1}{\sqrt{2^{n}}} \sum_{x=0}^{2^n-1} (-1)^{f(x)} H^{\otimes n}(|x\rangle)\\ 
\nonumber &= \frac{1}{2^n} \sum_{x=0}^{2^n-1} (-1)^{f(x)}  \sum_{j=0}^{2^n-1} (-1)^{x\cdot j} |j\rangle \\
\nonumber &=  \frac{1}{2^n} \sum_{j=0}^{2^n-1}|j\rangle \sum_{x=0}^{2^n-1} (-1)^{f(x)}  (-1)^{x\cdot j} 
\end{align}
Note the flipped sum in the final step.

\subsubsection{Step 4}

In the final step, we perform a measurement on the first $n$-qubits of the result from the previous step. If we measure these, the probability of measuring state $|j\rangle$ is 
\begin{equation}\nonumber
\left | \frac{1}{2^n} \sum_{x=0}^{2^n-1} (-1)^{f(x)}  (-1)^{x\cdot j} \right |^2
\end{equation}

In particular,  the probability of reading off state $|0\rangle^{\otimes n}$ is 

\begin{equation}\nonumber
\left | \frac{1}{2^n} \sum_{x=0}^{2^n-1} (-1)^{f(x)} \right |^2
\end{equation}

If $f(x)$ is constant, this probability is just $1$ - if $f(x)$ is balanced, this probability $0$. Hence, by looking at final measurement, we immediately know  if $f$ is balanced or not.

\subsection{Discussion}

While the Deutsch-Josza algorithm shows the advantages of quantum computing in a very specific setting, this problem/algorithm is not of practical use and mainly of interest to theoreticians who study algorithms. However, the solution to this problem does illustrate some general principles that are used to make Quantum algorithms powerful: making use of superposition to encode a lot of information compactly, and then using quantum operators to manipulate and extract the relevant information from them. It also lays the groundwork for more advanced algorithms that use these other principles (like making use of the probabilistic nature of the quantum computing) to solve more relevant complex problems.

One last point we mention here is that entanglement does not play any role in the above algorithm. However, as we mentioned above entanglement plays a fundamental role in advanced quantum algorithms. Even for the Deutsch-Josza algorithm, entanglement is necessary to implement the algorithm in practice \cite{arvind:2001}.

\section{ Next Steps}

If this article has whetted your appetite, and you want to learn more about quantum computing, and in particular more complex and relevant practical algorithms (like Shor's algorithm), then the next step should be to study an introductory quantum computing textbook. At this point there are many such books out there, and we will mention a few. One book that I have used and found to be useful is \cite{RP2014}. The book \cite{NM2008} is targeted toward computing professionals while \cite{W2022} even makes fewer assumptions about the background of the reader. \cite{NC2010} is a standard text on quantum computing and quantum information, but much more rigorous and aimed at advanced readers. Finally, \cite{SP2021} tackles the intersection of quantum computing and machine learning.

If you want to learn more about the overall industry context of quantum computing, the set of \href{https://www.digicatapult.org.uk/expertise/programmes/programme/quantum-computing/}{articles written at Digital Catapult} should be very useful.

If you want to be more hands on with quantum computing tools and algorithms, the \href{https://qiskit.org/}{Qiskit library} will be a good place to start. 

Finally, there are many online courses available for getting started on quantum computing. We recommend having a look around to see which best suits your needs.

We hope that you have found this review valuable. If you have any questions or feedback please contact the quantum team at Digital Catapult \href{mailto:quantum@digicatapult.org.uk}{\nolinkurl{quantum@digicatapult.org.uk}}.

\section{ Conclusion}

This article was an attempt to give a minimalist introduction to the most basic ideas of quantum computing using the quantum circuit model of computation. The introduction was minimal in the sense that it completely eschewed physics aspect of quantum computing, which can be counterintuitive and hard to grasp, and focused only on a mathematical/computational aspect, which can be understood with only some basic linear algebra and complex analysis under the belt. We also described one of the simpler quantum algorithms which demonstrates how quantum algorithms can be superior to classical ones and pointed the reader to references that they can use to learn more about quantum computing.

We hope that you have found this review valuable. If you have any questions or feedback please contact \href{quantum@digicatapult.org.uk}{the quantum team} at Digital Catapult.

\subsubsection*{Acknowledgements} We would like to thank Joseph Day-Evans for reviewing the paper and providing suggestions that have improved the paper.   

\begin{small}
  \bibliographystyle{plain} 
  \bibliography{quantum_minimal_introduction_arxiv}
\end{small}
\end{document}